\documentstyle[aps,pre,epsfig,preprint]{revtex}
\preprint{NAG39702}
\draft			

\begin{document}

\title{Aging in lattice-gas models with constrained dynamics}
			 
\author{Jorge Kurchan\cite{add1}}

\address{\'Ecole Normale Sup\'erieure de Lyon, 
46, All\'ee d'Italie, F-69364 Lyon Cedex 07, France.}

\author{Luca Peliti\cite{add2} and Mauro Sellitto\cite{add3}}

\address{Dipartimento di Scienze Fisiche and Unit\`a INFM, \\
Universit\`a ``Federico II'', Mostra d'Oltremare, Pad.~19,
I-80125 Napoli, Italy.}

\date\today

\maketitle

\begin{abstract}
We investigate the aging behavior of lattice-gas models
with constrained dynamics in which particle exchange 
with a reservoir is allowed. 
Such models provide a particularly simple interpretation of  aging
phenomena  as a slow approach to criticality.
They appear as the simplest three dimensional models exhibiting a
glassy behavior similar to that of mean field (low temperature
mode-coupling) models.

\end{abstract}

As the temperature of a supercooled liquid is reduced, its dynamics becomes
slower and slower and its viscosity increases. 
The glassy transition is usually located at the temperature $T_g$ where
the viscosity first exceeds the conventional value of $10^{13}$ Poise
\cite{Go}.
At lower temperatures, the glass is usually assumed to be ``solid'' in 
the sense that its molecules rattle within a ``cage'' formed by their
neighbors and do not move away from it, at least within experimental
time scales.
However the glassy state is not an equilibrium one since the evolution
of the system does not stop altoghether,
but rather keeps evolving at a slower and slower pace as the time 
$t_w$ elapsed since its quench increases.
This is the origin of the striking aging effects observed in glasses
\cite{St,Ho}. 
Aging corresponds to the property that, while one-time quantities like
the average energy, volume, etc., appear to be invariant under time 
translations, two-time correlations and responses exhibit a non 
trivial dependence on both of their arguments \cite{Saclay}.
Similar properties have also been observed in spin-glasses \cite{agingSG}. 
These aging properties have been thoroughly investigated within 
spin-glass models \cite{review} which are closely related to the 
mode-coupling theories of structural glasses \cite{Go}.
Aging appears provided one takes the infinite size limit
{\em before\/} taking the infinite time limit \cite{Cuku1}. 
In this
situation, the Boltzmann-Gibbs equilibrium distribution is
never reached. Conventional mode-coupling theories, which
are based on the hypothesis that the initial distribution
is the equilibrium one, are therefore inconsistent in this
limit below $T_g$.
 
The spin-glass models proposed as a description of 
structural glasses lack, however, a transparent physical 
interpretation in terms of particles and involve a complex, random 
hamiltonian which is hard to justify as a description of a fluid.
Their justification is rather {\it a posteriori} in the sense that the 
phenomena they exhibit recall the behavior of structural 
glasses.
On the other hand a class of very simple kinetic models 
have been introduced to describe the slowing down of the dynamics 
\cite{Fran,todos}.
These models are defined by kinetic rules involving a selection of
the possible configuration changes (``moves'') and are therefore 
called models with constrained dynamics.
The kinetic rules satisfy detailed balance and are 
compatible with a Boltzmann-Gibbs equilibrium distribution involving 
a hamiltonian, usually chosen to be a trivial one.
The constraints are alone responsible for the slowing down of the dynamics 
because, near any allowed configuration, there are only few configurations 
which satisfy them.

Some out of equilibrium properties in models of this kind have
been already discussed in \cite{Fori}.
We shall in this paper refer to the kinetic lattice-gas model
studied by Kob and Andersen (KA) \cite{Koan}.
The system consists of $N$ particles in a cubic  lattice of side $L$, 
($V=L^{3}$) with periodic boundary conditions.
There can be at most one particle per site. 
At each time step a particle and one of its neighbouring sites are 
chosen at random. 
The particle is moved if the three following conditions are all met:
(i) the neighbouring site is empty;
(ii) the particle has less than four nearest neighbours;
(iii) the particle will have less than four nearest neighbours after 
it is moved.
The rule is symmetric in time, detailed balance 
is satisfied and the allowed configurations have equal probability 
in equilibrium. 
In this model  the dynamics becomes slower and slower 
as the particle density $\rho$ increases.
Above a critical value of the density, $\rho_c$, particle diffusion
stops completely.
These investigations are carried on initializing the system in an 
``equilibrium'' configuration: therefore the approach cannot
be extended below the glass transition temperature without
becoming vulnerable to the same objection voiced above against 
the mode coupling theories:
how can one reach this ``equilibrium'' configuration?

We show in this paper, by numerical simulation, that it is possible to 
study aging 
in these models, if they are provided with a mechanism allowing one 
to realize a process analog to a quench.
We consider the three-dimensional KA model, but we allow the lattice 
gas to exchange particles on a single two-dimensional layer with a 
reservoir characterized by the chemical potential $\mu$ 
\cite{Alberto}. 
For each value of $\mu$ it is trivial to evaluate the equilibrium value
of the density  
$
\rho_{eq}(\mu)
$. 
If $\rho_{eq}(\mu)<\rho_c$, the system rapidly reaches the equilibrium 
state. There is therefore a critical value, $\mu_c$, of $\mu$, defined
by $\rho_{eq}(\mu_c)=\rho_c$. 
A quenching process corresponds to performing a jump of  $\mu$
from below to above  $\mu_c$.
Therefore, $\mu$ plays a role analogous to the
inverse temperature in mode-coupling theories.
We then observe that $\rho$ never exceeds $\rho_c$, but rather 
approaches it like a power law in time, and that the square 
displacement and the self-correlation function of the particle
violate time-translation invariance, since they explicitly depend on 
the waiting time, $t_w$, after the quench.

One can also perform ``annealing'' (compression) experiments, where 
$\mu$ is increased at a fixed rate.
In close resemblance with the behavior of real glasses, we observe
that the limit density depends on the compression rate, and tends to
$\rho_c$ as the rate decreases to zero.
We can then evaluate, by integration, the entropy variation  of the 
reservoir. The curves remain consistently above the equilibrium curve
$S=S_{eq}(\mu)$, and seem to reach a critical value, $S_c$, as the 
compression rate decreases to zero. This is reminiscent of the 
Kauzmann paradox, which is usually considered as an argument for the 
existence of a thermodynamical phase transition related to the glass 
one.
However in this case, we know that there is no such transition.

Hysteresis effects also appear when the reservoir's chemical potential
is varied cyclically.

As mentioned in the Introduction, KA studied the diffusion and the decay 
of the correlation functions evolving at fixed number of particles starting 
from a random (equilibrium) configuration.
They found that correlations decayed to zero with a characteristic time 
$\tau(\rho)$ that appears to diverge at $\rho_c \sim 0.881$, where also 
the diffusivity constant goes to zero.
For densities above $\rho_c$, the correlations  did not decay at all.

We  modify the KA model  by creating and destroying particles on a
single slice, say $(x,y,z)=(1,y,z)$ (the ``surface'') with the following 
Montecarlo rule: We choose a site on that slice at random. 
If it is empty we add a particle, if it is occupied we remove the particle
with probability ${\mbox{e}}^{-\mu}$. 
We alternate such sweeps of creation/destruction
with the ordinary diffusion sweeps. Since the creation/destruction mechanism
is very fast, the surface itself is always in equilibrium.
										
It is easy to calculate the thermodynamical equilibrium quantities
(we set the temperature $k_BT=1$ throughout):
$
1/\rho = 1+{\mbox{e}}^{-\mu} 
$,
where $\rho$ is the density. 
The entropy per unit volume is given by 
$				
S= - \rho \ln \rho  - (1-\rho) \ln (1-\rho)
$.

\begin{center}
\begin{figure}[f]
\epsfig{file=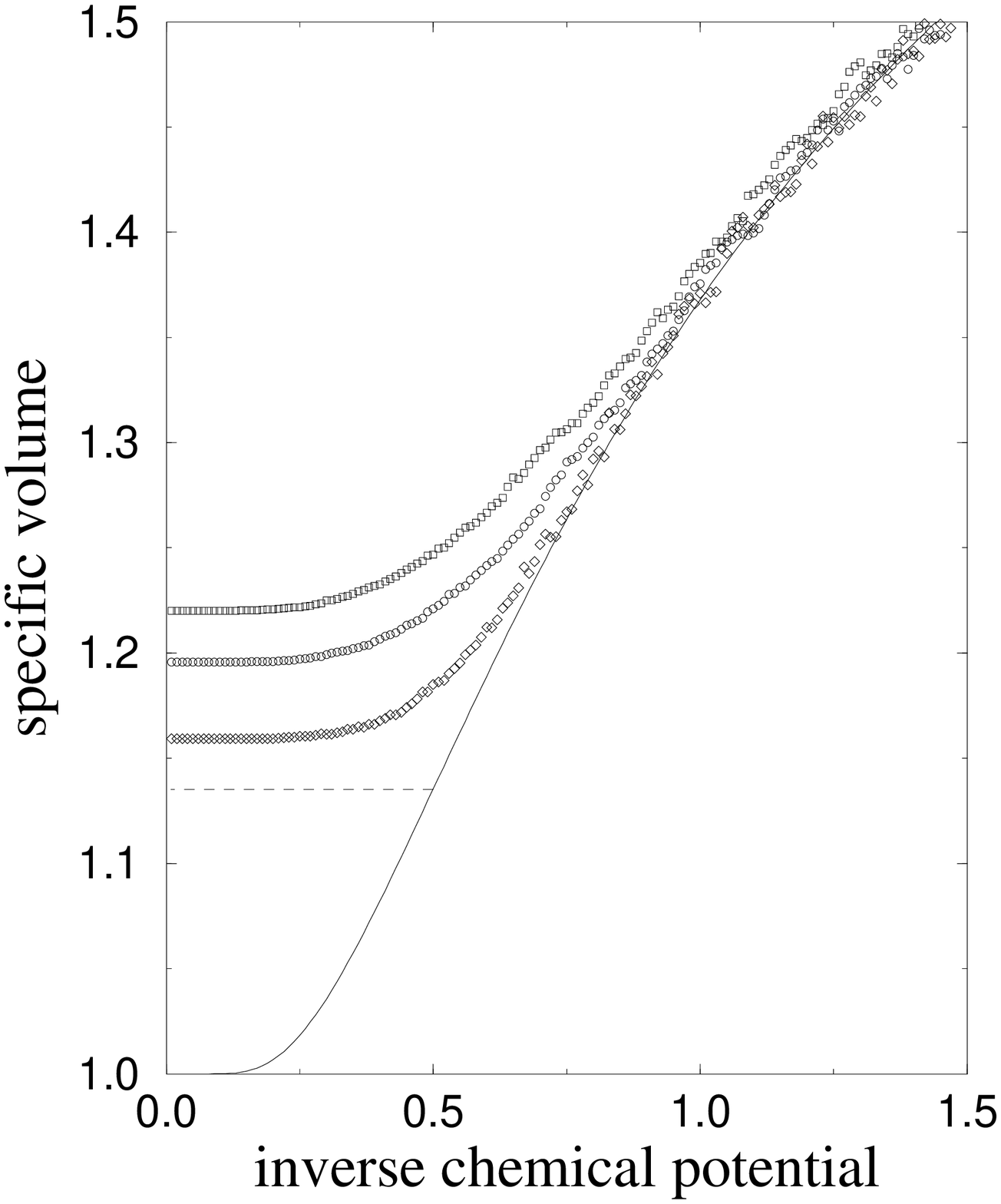,width=6cm,height=6cm} 
\hspace{2cm}
\epsfig{file=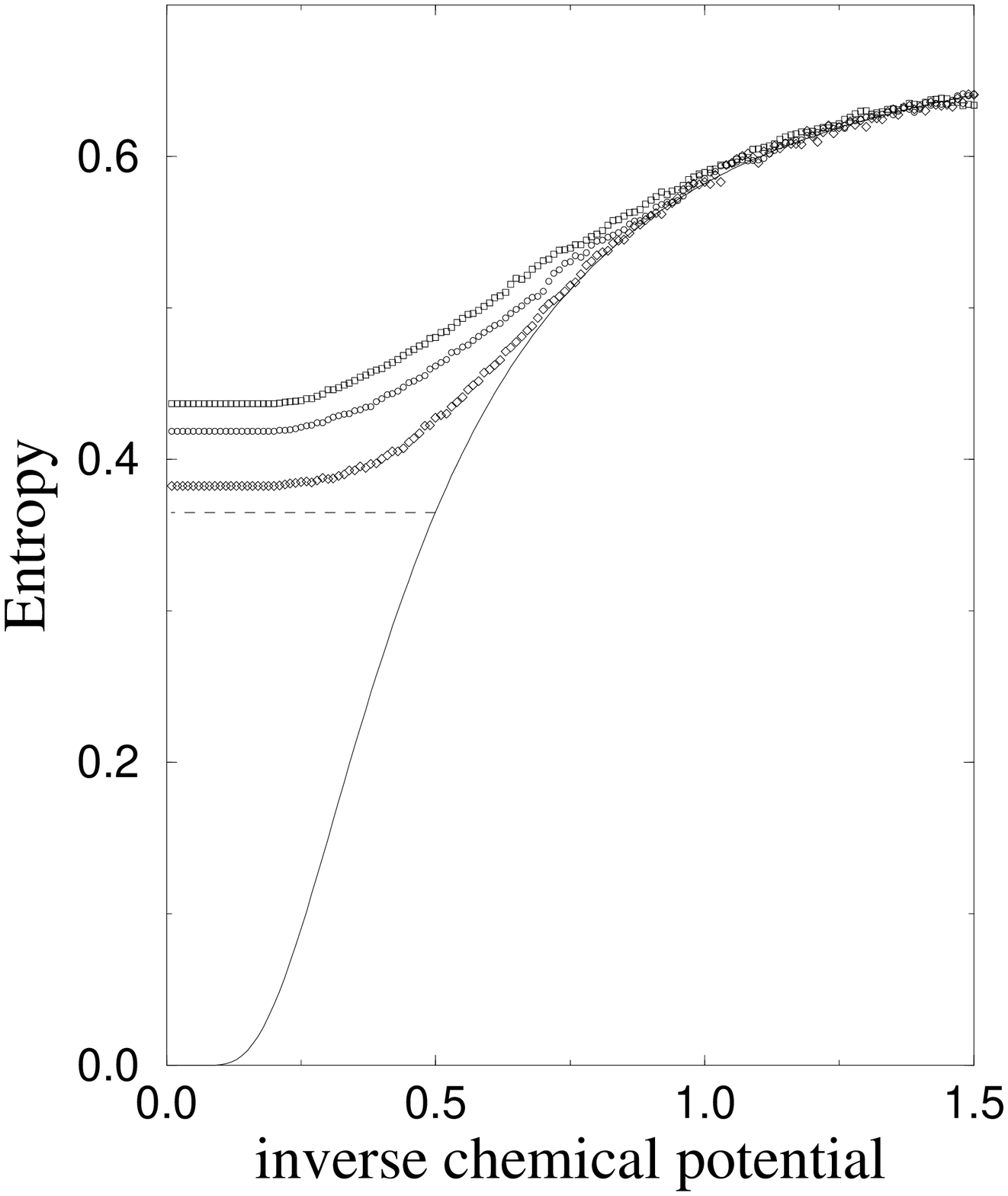,width=6cm,height=6cm} 
\vspace{1cm}
\caption{a: Compression experiment. The smooth curve is the equilibrium
state equation; while the dashed line is the KA critical value
of specific volume. 
b: `Entropy' as obtained by integration of compression experiment data.
The smooth curve is the equilibrium entropy. 
The dashed line is the KA critical value of entropy.
The compression rates are 
(from top to bottom) $3 \cdot 10^4, 10^5 ,10^6 $ sweeps per unit of 
inverse  $\mu$. System of size $20^3$, averages over five samples.  }
\label{annealing}
\end{figure}
\end{center}

We first consider compression experiments in which $1/\mu$ (a quantity 
analogous to temperature) is slowly lowered from a high  value to zero.
In Fig.~\ref{annealing}a we report the results of the specific volume 
$v=1/\rho$ versus $1/\mu$ for several annealing speeds. 
The smooth curve is the equilibrium state equation and the horizontal 
dashed line is the critical specific volume of KA  $v_c=1/\rho_c=1/0.881$. 
In Fig.~\ref{annealing}b, we show the  change in entropy of the reservoir, 
given by
$
S(\mu)= S(\mu_i) - \int_{\mu_i}^\mu  \mu d\rho
$.
The smooth curve is the equilibrium curve  $S$ vs. $1/\mu$, and 
the horizontal dashed  line is the entropy at the critical density.
These curves look quite similar to the energy vs. temperature and 
entropy versus temperature curves in the annealing of glasses.
As predicted, the system falls off equilibrium around $\mu_c$. 
We have checked that the bulk of the sample does not exhibit
density inhomogeneities except very close to the ``surface''.

\begin{center}
\begin{figure}[f]
\vspace{-3.3cm}
\epsfig{file=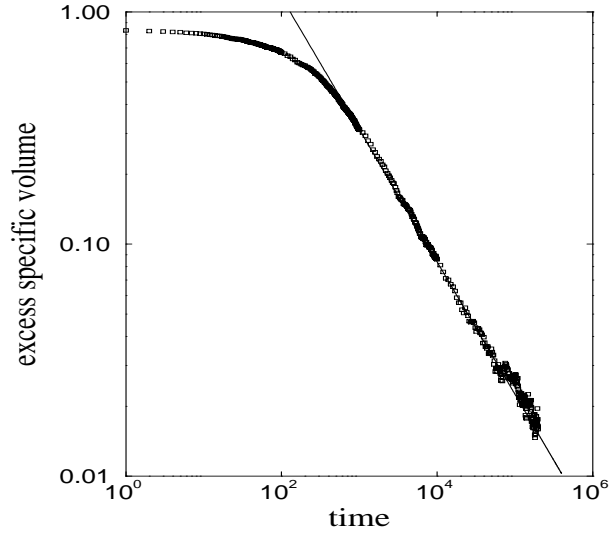,width=6cm,height=9.5cm} 
\vspace{1cm}
\caption{Relaxation of the excess specific volume $\Delta v=v(t)-v_{\infty}$
after a quench to the subcritical value $1/\mu= 1/2.2$.
The straight line corresponds to $\Delta v\propto t^{-0.57}.$}
\label{relax}
\end{figure}
\end{center}

We now turn to the behaviour of the system after a quench
to the subcritical value $1/\mu< 1/\mu_c$. Here we concentrate
on the behaviour of a large system at long but finite times. 
In Fig.~\ref{relax} we plot the specific volume minus its asymptotic 
value versus time after a quench to the subcritical value 
$1/\mu= 1/2.2$.
The data  can be fitted by a power law of the form  
$v(t) - v_{\infty} \sim  
t^{-\alpha}$, where $\alpha=0.57 \pm 0.02$. 
The asymptotic value of the density 
$ \rho_{\infty}=1/v_{\infty}=0.861$ is slightly lower than the KA value 
$\rho_c=0.881$, 
suggesting that the critical density might have a weak dependence on the 
way the particles entered the bulk  (although it might also be a finite-size 
effect).

\begin{center}
\begin{figure}[f]
\epsfig{file=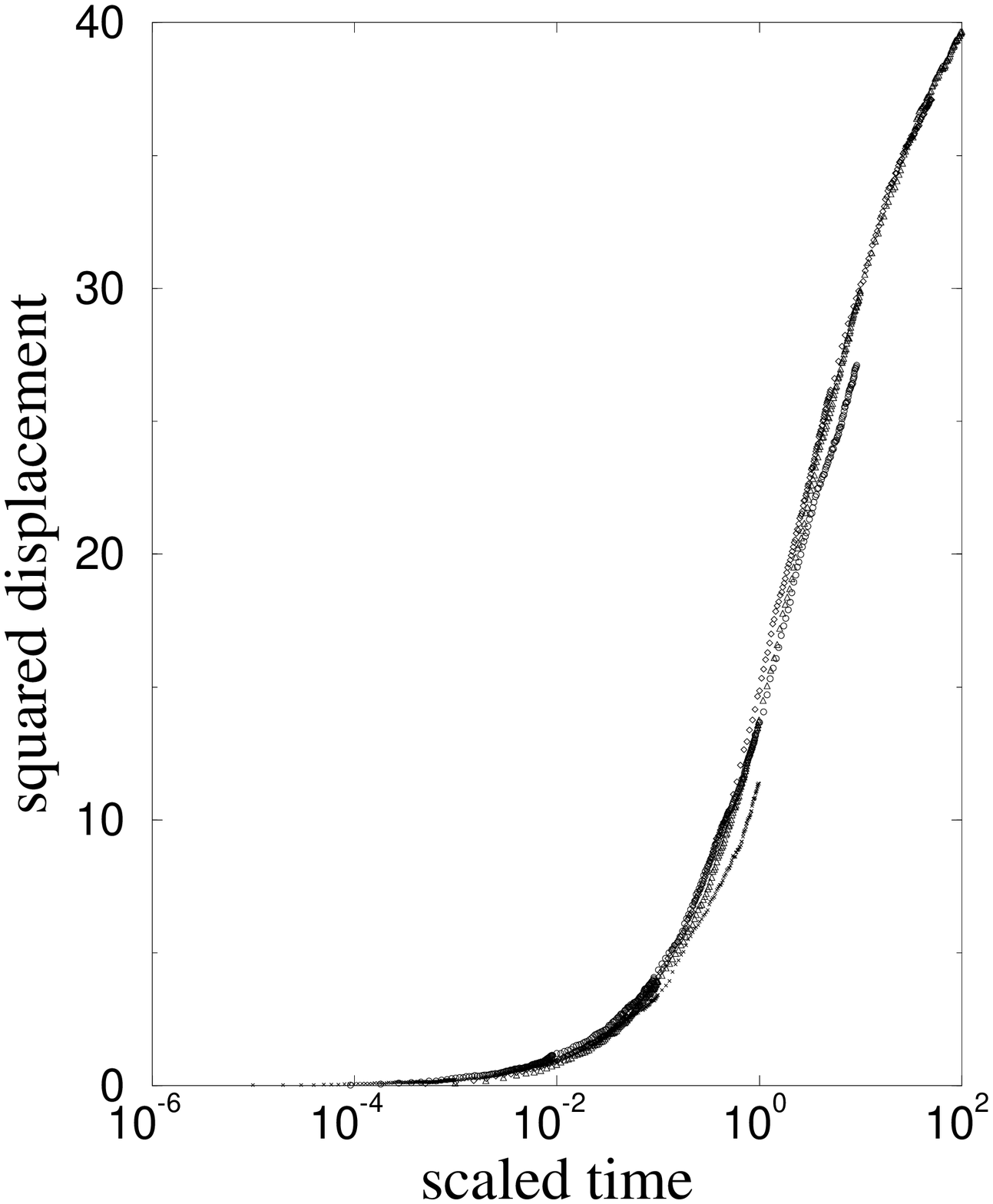,width=6cm,height=6cm} 
\vspace{1cm}
\caption{Squared displacement $R^2(t+t_w,t_w)$ vs. scaled time
$t/t_w^{eff}$ for $t_w=10,10^3,10^4,10^5$, and $t_0=10^3$
 ($t^{eff}_w=t_w+t_0$). 
System of size $20^3$, average over five samples.}
\label{deff}
\end{figure}
\end{center}

The mean square displacement $R^2(t+t_w,t_w)$ of a  particle between time
$t_w$ and time $t+t_w$ must be defined with some care, since the particles
may leave or enter the system.
We define it by averaging only over the particles which are present
at both times.
We find that diffusion indeed slows down as the waiting time
increases. 
However, the time $t(R^2)$ needed to reach a given square distance
is roughly proportional to $t_w$ only for long times.
In fact, we find here a phenomenon already encountered by Kob and Barrat
\cite{Bako} in Lennard-Jones systems: one has to consider an effective
waiting time that takes into account
the relaxation of the system that happened already before  the quench. 
Hence, one should define an effective
waiting time $t_w^{eff}=t_w + \tau_0$, where $\tau_0$ is the relaxation 
time characteristic of the  equilibrium situation before the quench. 
Fig.~\ref{deff} shows that, if one plots the squared displacement vs.\
$t/t_w^{eff}$, the curves lie roughly on top of each other. The 
deviations are probably due to finite size effects, e.g.
to particles that escape from the sample.
The plot exhibits the waiting-time dependence of the diffusion constant
and is consistent with a simple argument suggesting that diffusion is 
logarithmic in the presence of aging \cite{giorgio}.

One can also compute easily the effective potential of Franz 
and Parisi \cite{Frpa}.
This potential is a purely static quantity, and therefore is 
independent of the dynamical constraints.
Indeed, we have not found any signature for the dynamical transition in the 
effective potential.
However
the ``number of neighbours'' constraint is an idealization of microscopical  
(repulsive) energy terms, and as such it {\em should} be seen by the effective 
potential. 
It would be interesting to clarify this point further.

Since the size of the system considered here is finite, equilibrium     
must eventually be reached, provided that any two allowed configurations 
can be connected by a path of allowed moves.
It is possible to convince oneself that the kinetic rules allow an
initially empty lattice to be progressively filled in, leaving only     
O($1/L$) empty sites per unit volume, and that it is possible to 
find a path connecting almost any two allowed configurations, 
if necessary by letting the particles escape one by one by the way they 
got in.
Therefore, for any value of $1/\mu$, the equilibrium density will eventually 
be reached, but with times which diverge fast as 
$ L \rightarrow \infty$.
We have verified that a system of size $5^3$ equilibrates after a time
of order $10^8$ MC sweeps.
What this argument implies is that after a quench there is first the 
power-law approach to a critical density $\rho_c$, followed by a much 
slower (size-dependent) increase in density to the equilibrium density.

We have thus shown that  the out of equilibrium dynamics of
kinetic lattice-gas models reproduce the phenomenology
of more sophisticated glass models if one endows them with 
a surface mechanism of particle exchange with   a reservoir. 
Equivalently, we could have coupled the system to a piston and 
used the pressure as an external parameter instead of the reservoir's 
chemical potential.								
The {\em qualitative} behaviour of this model is similar to 
that of mean-field spin-glass models, whose dynamical equations 
are {\em exactly} (in the infinite-size limit) the ``idealised'' 
mode-coupling equations and their correct low-temperature extension.
The roles played by the inverse chemical potential and the inverse 
density (specific volume)  in the present model are analogous  to  
the ones played by temperature and energy respectively, in the 
mean-field case. 
When quenched below  $T_c$, mean-field spin-glass models never equilibrate. 
Their energy relaxes to a ``threshold'' value $E_{thres}(T)$ which is 
higher than the equilibrium one \cite{Cuku1}, and can be characterized
as the value below which the phase space breaks into disconnected
components.
In our case, the moves allowed by the kinetic constraints keep the phase
space of the system connected only for $\rho<\rho_c$.
Therefore the density approaches
this threshold value, and the corresponding manifold of allowed states
approaches to the incipient ``percolation'' cluster
characteristic of $\rho_c$ \cite{Campbell}.
Only for finite sizes the system is able to reach values of $\rho$ larger
than $\rho_c$, just as energy values below the threshold can only be 
reached by finite systems in mean-field spin-glasses \cite{Cuku1,Babume}.

In short: in both cases the system ages because it slowly evolves towards 
a situation in which  the available phase-space becomes completely 
disconnected \cite{Campbell}.
The closer the system to  the ``phase-space percolation'' threshold level,
the slower its dynamics becomes.
In this sense the system ages because it approaches criticality \cite{SOC}.

To summarize, lattice gas models with constrained dynamics, in spite of 
their simplicity, capture the essential features of the aging behavior 
of three dimensional glasses.

We are indebted to J.-L. Barrat for suggesting to look at the effective
potential and to J.-P. Bouchaud, L. Cugliandolo, 
P. Holdsworth and W. Kob for useful discussions and suggestions.
M.S. acknowledges the hospitality
of the Laboratoire de Physique Th\'eorique of \'Ecole Normale Sup\'erieure
de Lyon, where part of this work was performed.

\begin{references}


\bibitem{add1} E-mail address: {\tt Jorge.Kurchan@enslapp.ens-lyon.fr}


\bibitem{add2}
Associato INFN Sezione di Napoli.
E-mail addresses: {\tt luca@turner.pct.espci.fr, peliti@na.infn.it}

\bibitem{add3}  E-mail address: {\tt Mauro.Sellitto@na.infn.it  }

\bibitem{Go} See e.g.:
W. G\"otze, L. Sj\"ogren, {\it Rep. Prog. Phys.} {\bf 55}, 241 (1992).




\bibitem{St}
L.C.E. Struik, {\it Physical Aging in Amorphous Polymers and
Other Materials}, (Houston: Elsevier, 1978).

\bibitem{Ho}
I. Hodge, {\it J. Non Cryst. Solids} {\bf 169}, 211 (1994); 
{\it Science} {\bf 267}, 1945 (1996).

\bibitem{Saclay}
For a recent review, see:
E. Vincent, J. Hammann, M. Ocio, J.-P. Bouchaud and L.F. Cugliandolo,
in 
{\it Proceeding of the Sitges Conference on glassy systems},
June 1996, edited by E. Rub\'{\i}, cond-mat {\bf 9607224}.

\bibitem{agingSG}
L. Lundgren, P. Svedlindh, P. Nordblad and  O. Beckman,
{\it Phys. Rev. Lett.} {\bf 51} 911 (1983);
E. Vincent, J. Hammann and M. Ocio; in {\it Recent Progress in Random Magnets},
edited by D. H. Ryan, (Singapore: World Scientific, 1992) p. 207.

\bibitem{review} 
J.-P. Bouchaud, L.F. Cugliandolo, J. Kurchan and 
M. M\'ezard, cond-mat {\bf 9702070}, and references therein;
for a review on aging simulations in spin-glasses see:
H. Rieger, {\it Ann. Rev. of Comp. Phys.} II, ed.
D. Stauffer  (World Scientific, Singapore, 1995). 

\bibitem{Cuku1} 
 L.F. Cugliandolo and  J. Kurchan, {\it Phys. Rev. Lett.} {\bf 71},
 173 (1993); {\it Phil. Magaz. B} {\bf 71},  50 (1995).

\bibitem{Fran} G.H. Fredrickson and H.C. Andersen, {\it Phys. Rev. Lett.}
{\bf 53}, 1244 (1984); {\it J. Chem. Phys. } {\bf 83}, 5822 (1985);
 G.H. Fredrickson and S.A. Brawer, {\it J. Chem. Phys.} 
{\bf 84}, 3351 (1986). 

\bibitem{todos}  
W. Ertel, K. Frob\"ose and J. J\"ackle, {\it  J. Chem. Phys.} {\bf 88}, 
5027 (1988). 
K. Frob\"ose, {\it J. Stat. Phys.} {\bf 55}, 1285 (1989). 
J. J\"ackle, K. Frob\"ose and D. Kn\"older, {\it  J. Stat. Phys.} {\bf 63},
249 (1991). 
J. Reiter, F. Mauch and J. J\"ackle, {\it Physica A} {\bf 184}, 458 (1992). 
J. J\"ackle and D. Sappelt, {\it Physica A} {\bf 192}, 691 (1993).

\bibitem{Fori} E. Follana and F. Ritort, {\it Phys. Rev. B} {\bf 54}, 
930 (1996).   See also a density functional lattice model in: 
F.G. Padilla and F.~Ritort, cond-mat {\bf 9703095} (1997),
 where creation and destruction of particles within the bulk is allowed.

\bibitem{Koan} W. Kob and H.C. Andersen, {\it Phys. Rev. E} 
{\bf 48},  4364 (1993).

\bibitem{Alberto} Coupling a kinetic lattice-gas model with a particle
reservoir has been considered by A. Imparato, University of Naples thesis,
unpublished. 
D. Stariolo, cond-mat {\bf 9612082}, 
has suggested that aging may be a consequence
of mass non-conservation in diffusion models.

\bibitem{Frpa} S. Franz and G. Parisi, {\it J. Phys. (France) I} {\bf 5}, 
1401 (1995).

\bibitem{Bako}		
W. Kob and J.-L. Barrat, cond-mat {\bf 9704006}.

\bibitem{giorgio} G. Parisi, cond-mat {\bf 9703219}.

\bibitem{Babume}  A. Barrat, R. Burioni and  M. M\'ezard,
{\it J. Phys. A} {\bf 29}, L81 (1996).

\bibitem{Campbell} For discussions of this kind, see 
I.A. Campbell, {\it J. Phys. Lett.} {\bf 46}, L1159 (1985);
 {\it Phys. Rev. B}
{\bf 33}, 3587 (1986). 
 C.M. Newman and D.L. Stein, {\it Phys. Rev. Lett. } {\bf 72}, 2286
(1994);
D.L. Stein and C.M. Newman, {\it Phys. Rev. E} {\bf 51}, 5228 (1995).

\bibitem{SOC} The connection between aging and self-organized criticality
has been explored by:
S. Boettcher and M. Paczuski, cond-mat {\bf 9702054}; see also
C. Tang and P. Bak, {\it Phys. Rev. Lett. } {\bf 60}, 2347
(1988). 
We thank P. Bak for have pointed out to us these references.


\end{references}
\end{document}